\def\ln{\ell{n}}
\documentstyle[12pt]{article}

 {\parindent 0.5cm \textwidth 15.0cm \textheight
23.5cm \hoffset=-0.8cm \voffset=-3cm
  \let\LARGE=\large
 \let\large=\normalsize
\begin{document}
\begin{titlepage} \vspace{0.2in} \begin{flushright}
MITH-93/16 \\ \end{flushright} \vspace*{1.5cm}
\begin{center} {\LARGE \bf  The Mass Ratio Between Neutrinos and
Charged Leptons\\} \vspace*{0.8cm}
{\bf Giuliano Preparata$^{a,b}$ and She-Sheng Xue$^{a,\dagger}$\\}
a) INFN, Section of Milan, Via Celoria 16, Milan, Italy\\
b) Physics Department, University of Milan, Italy\\
\vspace*{1.8cm}
{\bf   Abstract  \\ } \end{center} \indent

In the framework of the recently proposed electroweak theory on a Planck
lattice, we are able to solve approximately the lattice Dyson equation for
the fermion self-energy functions and show that the large difference of charged
lepton and neutrino masses is caused by their very different gauge couplings.
The predicted mass ratio ($10^{-5}\sim 10^{-6}$) between neutrinos and charged
lepton is fully compatible with present experiments.
\vfill \begin{flushleft} 1st June, 1993   \vspace*{3cm} \\
\noindent{\rule[-.3cm]{5cm}{.02cm}} \\
\vspace*{0.2cm} \hspace*{0.5cm} ${}^{\dagger}$
E-mail address: xue@milano.infn.it\end{flushleft} \end{titlepage}

Since their appearance neutrinos have always been extremely peculiar. In the
sixty years of their life, their charge neutrality, their apparent
masslessness, their left-handedness have been at the centre of a conceptual
elaboration and an intensive experimental analysis that have played a major
role in donating to mankind the beauty of the electroweak theory. V-A theory
and Fermi universality would possibly have eluded us for a long time had the
eccentric properties of neutrinos, all tied to their apparent masslessness, not
captured the imagination of generations of experimentalists and theorists
alike.

However, with the consolidation of the Standard Model (SM) and in particular
with the general views on (spontaneous?) mass generation in the SM, the
observed (almost) masslessness of the three neutrinos
($\nu_e,\nu_\mu,\nu_\tau$) has recently come to be viewed as a very problematic
and bizarre feature of the mechanism(s) that must be at work to produce the
very rich mass spectrum of the fundamental fields of the SM. Indeed, in the
(somewhat worrying) proliferation of the Yukawa couplings of fermions to the
Higgs fields that characterizes the generally accepted SM, no natural reason
can be found why the charge-neutral neutrinos are the fundamental particles of
lowest mass; for in the generally accepted minimal Higgs mechanism, the actual
values of the fermion masses are in direct relation with the strengths of their
couplings to the Higgs doublet, and it appears rather bizarre that nature has
chosen to create a very sophisticated mass pattern by the mere fine-tuning of a
large number of parameters.

On the other hand we believe that it cannot be a simple dynamical accident that
the only charge-neutral fermions that populate the SM are also the lightest. In
this letter we wish to show that this most remarkable property of Nature can
naturally be understood in the framework of the Planck Lattice Standard Model
(PLSM) that we have recently introduced \cite{xue91,xue92,xue93}, based on the
general hypothesis that, as a result of the violent quantum fluctuations of the
metrical (gravitational) field, at the Planck length $a_p\simeq 10^{-33}$ cm,
space-time is no more a 4-dimensional continuum, but can be well represented by
a (random) lattice with (average) lattice constant equal to
$a=a_p$\cite{planck}. Here we would only like to stress that in order to avoid
a well known theorem by Nielsen and Ninomiya \cite{nogo}, which prevents a
simple transcription of the usual continuum SM lagrangian to the lattice, we
have been forced to add new 4-fermi terms of the Nambu-Jona Lasinio type
(NJL)\cite{nambu}. Recently, we analyzed the solutions of the Dyson equations
involving the NJL interactions only, with the following results\cite{xue92}:
\begin{enumerate}
\item
the fundamental chiral symmetry of the full Lagrangian is
spontaneously broken, and as a consequence \underbar{only one family} in
the quark sector,
which is identified with the top quark and bottom quark doublet,
acquires a mass. The lepton sector remains massless owing to its different
colour structure.

\item
the consistent solution of the gap equations also produces non-zero
Wilson-type parameters $r_q$ and $r_l$ and mass counterterms for each quark
and lepton.
\end{enumerate}
The effective action that is left after such massive rearrangement of the
vacuum is ($a=a_p$)
\begin{eqnarray}
S & = & S_G + S_D + \sum_{xF} \bar \psi (x) m_F \psi (x)\label{action}\\
&& - {1 \over 2a} \sum_{Fx \mu} \bigl[ \bar \psi^F (x) \left(L^F_\mu(x) +
R^F_\mu(x )\right) r_F U^F_\mu (x) \psi^F (x + a_\mu) + {\rm h.c.} \bigr]
+\cdot\cdot\cdot,
\nonumber
\end{eqnarray}
where $S_G$ is the usual Wilson gauge action, $S_D$ the usual Dirac action,
$F=l (q)$ denotes the lepton (quark) sector, $m_F, r_F$ are matrices in
flavour and weak isospin space, and the dots denote the NJL
interaction and the necessary
counterterms. Gauge links are given by
\begin{equation}
L^F_\mu(x) = U^L_\mu(x) V^{Y^F_L}_\mu(x);\;\;R^F_\mu(x) =
\pmatrix{
V^{Y^F_{R_1}}_\mu (x) &         0           \cr
0               &   V_\mu^{Y^F_{R_2}} (x)   \cr} ,
\label{l}
\end{equation}
where $
U^q_\mu(x) \in SU_c(3), \;\;U^L_\mu(x) \in SU_L(2) \; {\rm and }\; V_\mu(x)
\in U_Y(1)$.
Turning on gauge interactions we approximately analyzed the Dyson equation for
the top and bottom doublet and obtained a mass ratio between the top and bottom
quarks \cite{tb} in agreement with recent experimental indications.
In this note, we apply an analogous analysis to the lepton sector. Our aim
is, of course, to see whether gauge interactions are capable of yielding the
large observed differences between the masses of the charged leptons and
the neutrinos.

When we take into consideration in addition to the NJL interaction the gauge
interactions
of the action (\ref{action}), the
Dyson equations for the lepton sector have the structure depicted
diagrammatically in Fig.1, which can be written as:
\begin{eqnarray}
\Sigma^Q_c(p) &= &m_Q\sigma+\sum_g C^Q_g(p);\label{gap}\\
\sigma &=& {2g_1\over N_c} \int ^\pi _{-\pi} {d^4q \over (2 \pi)^4} \; {1 \over
\sin^2
q_\mu + [m_Q a + r_lw(q)]^2}+\cdot\cdot\cdot,
\label{njl}
\end{eqnarray}
where $\sum^Q_c(p)\, (\Sigma_c^Q(0)=m_Q)$ are the leptons' self-energy
functions; the second term in the rhs of (\ref{gap}) denotes the contributions
of the relevant
gauge interactions and $\sigma$ stands for the contributions (dots for
high-order contributions) of the NJL interactions, which are supposed to be the
same for neutrinos ($Q=0$) and charged leptons ($Q=-1$) because $m_Qa\simeq 0$,
$r_Q=r_l\;(Q=0,-1)$. We remark that the terms that diverge like ${1\over a}$
have
all been consistently cancelled by mass counterterms \cite{xue92}.

For external momenta $p_\mu a \ll 1$, we divide the integration domain over the
variable $q_\mu$ into two regions: the ``continuum" region: $0 \leq |q_\mu a|
\leq \epsilon$ and the ``lattice" region: $\epsilon \leq |q_\mu a| \leq \pi$,
where $|ap_\mu| \ll \epsilon \ll \pi$. With this separation we can write the
contribution of the gauge interactions as:
\begin{eqnarray}
C^Q_g(p)& =&\tilde C^Q_g(p,\Lambda)+\theta^Q_g(\epsilon,r_l)
+\delta^Q_g(\epsilon,r_l),\label{separation}\\
\tilde C^Q_g(p,\Lambda) &=& {1\over 4\pi^2}\int_\Lambda d^4q\, {\lambda_g(q^2)
\over [(p-q)^2+m_g^2]}\,{\Sigma^Q_c(q)\over [q^2+m^2_Q]},
\label{c}
\end{eqnarray}
where the continuum-region integral is up to $\Lambda=\epsilon\Lambda_p$,
$(\Lambda_p ={\pi\over a}\simeq 10^{19}$GeV). As for the
$\ln\epsilon$-divergent
contribution in the lattice-region, one has
\begin{eqnarray}
\theta^Q_g(\epsilon,\Lambda) &=&\int_{[\epsilon,\pi]}{d^4l\over 16\pi^2}\,{
\lambda^Q_g(l)\Sigma^Q_c(l)\over \sin^2{l_\mu\over2}}\;{\cos^2{l_\mu\over2}
\over (\sin^2(l)+(m_Qa+rw(l))^2)}\nonumber\\
&&\simeq \lambda^Q_g(\Lambda_p)\,m_Q\,(\bar c_2(r_l)-{1\over2}\ln\epsilon),
\label{theta}
\end{eqnarray}
where we take the asymptotic mean-field value $\Sigma^Q_c(|l|\geq\epsilon)
\simeq m_Q$ and numerically calculate $\bar c_2(r_l)$, which is plotted
in Fig.2. Note that the $q^2$-dependent renormalized couplings
are given by
\begin{equation}
\lambda^Q_g(q^2)=\lambda^Q_g(m_Q^2)Z_3(q^2,m_Q^2);\hskip2cm\lambda^Q_g(m_Q^2)
={3\over\pi}{\beta^Q_L\beta^Q_R(m_Q^2)\over 4\pi}
\label{coupling}
\end{equation}
where $(\beta^Q_{L,R})(m)$ are left- and right-handed gauge couplings
to different fermions ($Q$) on the mass-shell $m=m_Q$,
and $Z_3(q^2,m^2_Q)$ are normal gauge field renormalization functions in the
abelian and non-abelian cases ($SU_L(2)$). In eq.~(\ref{c})
the regular contributions from the lattice-region are
\begin{eqnarray}
\delta_{A,Z}^Q(\epsilon,r_l)\!
&\!=\!&\!-\Big({3\over\pi}{(\beta^Q_L+\beta^Q_R)^2(\Lambda_p)\over 4
\pi}\Big)r_l^2\int_{[\epsilon,\pi]}{d^4l\over16\pi^2}{\Sigma_c^Q(l)\over
(\sin^2(l)+(m_Qa+rw(l))^2)}\label{daz}\\
\delta^{(0,-1)}_{W^\pm}(\epsilon,r_l)_j\!
&\!=\!&\!-\Big({3\over\pi}{(\beta^Q_L\!+\!\beta^Q_R)^2(\Lambda_p)\over 4
\pi}\Big)r_l^2\!\int_{[\epsilon,\pi]}\!{d^4l\over16\pi^2}{\sum_{i=1}^3
|V_{KM}^{ji}|^2 \Sigma_c^{(-1,0)}(l)_i\over
(\sin^2(l)+(m_{(-1,0)}a+rw(l))^2)},
\label{dw}
\end{eqnarray}
where $V_{KM}$ is a KM-type matrix for lepton sector. Since (\ref{daz})and
(\ref{dw}) have no $\ln\epsilon$ divergence, we may safely take the limit
$\epsilon\rightarrow 0$.

Due to the fact that there are extra terms arising from $\bar
c_2(r_l)$ [eq.(\ref{theta})], $\delta^Q_g(r_l)$ [eqs.(\ref{daz},\ref{dw})]
and NJL terms in (\ref{gap}), the gap-equation (\ref{gap}) can have
consistent massive solution
($m_Q\not=0$) even for small gauge couplings\cite{bardeen}. In this preliminary
discussion, we take a trivial (unity) KM matrix in (\ref{dw}) and the Landau
mean-field approximation $\Sigma^Q_c(q)\simeq m_Q$ in
(\ref{c},\ref{theta},\ref{daz},\ref{dw}). Thus, for each lepton family,
we have from (\ref{gap})
\begin{eqnarray}
m_{\nu} &= & m_\nu(NJL)+m_{\nu}\delta^\nu_Z(r_l)+m_l\delta^\nu_W(r_l)
\label{n}\\
m_l &=& m_l(NJL)+m_l\Big(\bar C_A^l+\bar C_Z^l\Big)+m_l\Big(\delta^l_A(r_l)
+\delta^l_Z(r_l)\Big)+m_\nu\delta^\nu_W(r_l),
\label{meaneq}
\end{eqnarray}
where we can see that the gauge field contributions $ C^Q_g(p)$ in
eq.~(\ref{c})
play the essential role in distinguishing between neutrinos and charged
leptons.
We notice that for neutrinos $\tilde C_g^{(0)}=\theta^{(0)}_g=0 (g=A_{em},
W^\pm$ and $Z^\circ)$, while for charged leptons $\tilde C_{W^\pm}^{(-1)}=
\theta_{W^\pm}^{(-1)}=0$. As for the gap-equation of charged leptons
(\ref{meaneq}), we have
\begin{equation}
\bar C^Q_g=\tilde C^Q_g+\theta^Q_g={1\over 4\pi^2}\int_{\Lambda_p} d^4q
{\lambda_g(q^2)
\over [(p-q)^2+m_g^2]}{1\over [q^2+m^2_Q]}+\lambda^Q_g(\Lambda_p)
\bar c_2(r_l),
\label{bar}
\end{equation}
where the arbitrary $\epsilon$ scale disappears, as it should,
due to the cancellation of the
$\ln\epsilon$ term between $\tilde C^Q_g(p,\Lambda)$ [eq.(\ref{c})]
and $\theta^Q_g(\epsilon, r_l)$ [eq.(\ref{theta})]. By using the
renormalization
group relations (\ref{coupling}) to estimate the values of $\beta_L(\Lambda_p),
\;\beta_R(\Lambda_p)$ $(\alpha(\Lambda_p)\simeq 1.56\alpha(m_e)$,
$\sin^2\theta_w(\Lambda_p)\simeq 2.32\sin^2\theta_w(m_Z^2))$ in
(\ref{theta},\ref{daz},\ref{dw},\ref{bar})
$\left[\alpha\equiv\alpha(m_e)={1\over 137},\, \sin^2\theta_w\equiv
\sin^2\theta_w(m_Z^2)=0.23\right]$, one gets ($\bar C_g^l=C_g^{(-1)}$)
\begin{eqnarray}
\bar C_A^l &\simeq &\Big[3\cdot 10^{-3}\Big(\ln{\Lambda_p\over m_l}\Big)^2
+0.75\,\ln{\Lambda_p\over m_l}+1.5\,\bar c_2(r_l)\Big]\alpha;\nonumber\\
\bar C_Z^l &\simeq &\Big[3.6\cdot 10^{-3}\Big(\ln{\Lambda_p\over m_Z}\Big)^2
+0.37\,\ln{\Lambda_p\over m_Z}+0.18\,\bar c_2(r_l)\Big]
\Big({\sin^2\theta_w\over 4\pi}\Big)\nonumber\\
&&-\Big[1.2\cdot 10^{-3}\Big(\ln{\Lambda_p\over m_Z}\Big)^2
+0.37\,\ln{\Lambda_p\over m_Z}+0.18\,\bar c_2(r_l)\Big]\alpha,
\label{2delta}
\end{eqnarray}
where $m_Z=91.2$GeV. It will be seen soon that $\delta^Q_g(r_l)$,
[eqs.(\ref{daz},\ref{dw})], are small numbers ($\sim 10^{-5}$) due to the
smallness of gauge
couplings and of $r_l^2$. Neglecting $\delta^Q_g(r_l)$ in (\ref{meaneq}) and
adopting the four-fermi coupling $g_1$ determined by the top quark mass, one
can show that the gap equation for charged leptons (\ref{meaneq}) has a
consistent solution $m_l\sim O$(MeV)$ \ll \Lambda_p$ for small gauge couplings
\cite{xue93}. However, for the time being, we are not in a position to obtain
the correct spectrum of charged leptons \cite{xuewill}. As for the
gap equation for neutrinos (\ref{n}), the consistent solution must be
$m_\nu\ll m_l$ because $(1-\sigma)\sim O(\alpha\ln{\Lambda_p\over m_l})\gg
\delta^{(0)}_g(r_l)$. In fact, in eqs.~(\ref{n}) and (\ref{meaneq}),
we have seen the emergence of the hierarchy spectrum of neutrinos and
charged leptons due to their very different gauge couplings.

Although we are still not able to calculate the masses of neutrinos and charged
leptons based on eqs.~(\ref{n}) and (\ref{meaneq}), their mass ratio
can be estimated. It is easy to see that the solution to (\ref{meaneq}) is
simply
\begin{eqnarray}
&&\delta^\nu_W(r_l)m_l^2-2\Delta_lm_lm_\nu-\delta^l_W(r_l)m_\nu^2=0\nonumber\\
&&2\Delta_l=-\Big(\bar C_A^l+\bar C_Z^l\Big)-\delta^l_A(r_l)
+\Big(\delta^\nu_Z(r_l)-\delta^l_Z(r_l)\Big).
\label{fine}
\end{eqnarray}
{}From the definitions (\ref{daz}), one obtains
\begin{eqnarray}
\delta^\nu_Z(r_l)-\delta^l_Z(r_l) &\simeq & 82.2\,\alpha\, r^2_l\,G(r_l);
\label{lldelta}\\
\delta^l_A(r_l) &\simeq &58.9\,\alpha\, r_l^2\, G(r_l),\label{ldelta}\\
G(r_l)&=&\int{d^4p\over (2\pi)^4}{1\over \sin^2(l_\mu)+(m^Qa+r_lw(l))^2}
\label{gr}
\end{eqnarray}
where $G(r_l)$ as a function of $r_l$ is plotted in
Fig.3, while for $\delta^Q_W(r_l)$ [eq.(\ref{dw})], where
$\beta_R=0,\beta_L={g_2\over \sqrt{2}}$ and $g_2$ is the $SU_L(2)$ gauge
coupling, we have
\begin{equation}
\delta_W^l(r_l)=\delta_W^\nu(r_l)=-{3\pi\over 2}{\alpha(\Lambda_p)\over\sin^2
\theta_w(\Lambda_p)}r_l^2G(r_l)\simeq-13.72\,\alpha\,r^2_l\,G(r_l).
\label{ndelta}
\end{equation}
Thus, solving (\ref{fine}), we obtain
\begin{equation}
m_\nu={\Delta_l+\sqrt{(\Delta_l)^2+(\delta^l_W(r_l))^2}\over\delta^l_W(r_l)}
m_l.
\label{result}
\end{equation}
We see that $\Delta_l< 0$ and $\delta_W\ll 1$ are crucial for obtaining
$m_\nu\ll m_l$.
For $r_l\simeq 0.01\sim 0.04$\footnote{ In\cite{xue92} we show that the
Wilson parameter for quarks ($r_q=0.28\sim 0.3$) is determined by minimizing
vacuum energy, yielding for the Wilson parameter for leptons
$r_l=0.01\sim 0.04$.},one gets
$\bar c_2(r_l)=0.65\sim 0.62$ and
$G(r_l)=0.62\sim 0.6$, which leads approximately to
\begin{equation}
m_{\nu_l} \simeq (1.0\cdot 10^{-5}\sim 1.6\cdot 10^{-4})m_l.
\label{ratio}
\end{equation}
Putting the experimental charged lepton masses ($m_e=0.5$MeV,
$m_\mu=105$MeV, $m_\tau=1774$MeV) into (\ref{ratio}), we
predict the neutrino masses:
\begin{eqnarray}
m_{\nu_e} &\simeq (5\sim 128) {\rm eV},\nonumber\\
m_{\nu_\mu} &\simeq (1\sim 26) {\rm keV},\nonumber\\
m_{\nu_\tau} &\simeq (17\sim 284){\rm keV},
\label{prediction}
\end{eqnarray}
which are compatible with present experiments, even though our analysis
is still at a rather rudimentary level.
Leaving aside the uncertainties of our analysis, which may well exceed
the theoretical uncertainty \footnote{Note that the main uncertainty comes from
the uncertainty of
the Wilson parameter $r_l$.} appearing in our predictions, we believe that we
can say with a certain degree of confidence that:
\begin{enumerate}
\item
eqs.~(\ref{prediction}) represent the first (successful?) attempt to derive
within a complete closed dynamical scheme the mass ratios between the
members of the lepton doublets;
\item
these mass ratios are, to the approximation we work in, universal;
\item
the large difference in (\ref{prediction}) between the masses of the neutral
and charged members of the lepton weak doublets
finds its natural and convincing origin in the their gauge-couplings,
and in particular in the neutrino lack of electromagnetic interactions,
which most likely is at the root of the mechanism by which charged leptons
acquire their masses\cite{xuewill};
\item
the crucial role played by the Wilson parameter $r_l$ in yielding a
connection between the masses of the charged leptons and their neutrinos,
through the high
energy($\sim\Lambda_p$) coupling of the right-handed neutrinos to their
left-handed counterparts via the charged leptons and the $W^\pm$-bosons
(\ref{action}), should be viewed as a clearly unique dynamical feature of our
proposal.

\end{enumerate}

\newpage  \pagestyle{empty}
\begin{center} \section*{Figure Captions} \end{center}

\vspace*{1cm}

\noindent {\bf Figure 1}: \hspace*{0.5cm}
The diagrammatic form of the Dyson equation for
the lepton sector.

\noindent {\bf Figure 2}: \hspace*{0.5cm}
The function $\bar c_2(r_l)$ in terms of $r_l$.

\noindent {\bf Figure 3}: \hspace*{0.5cm}
The function $G(r)$ in terms of $r_l$.

}

\begin{thebibliography}{99}

\bibitem{xue91}
G.~Preparata and S.-S.~Xue, {\sl Phys.~Lett.} {\bf B264} (1991) 35;
{\sl Nucl.~Phys.} {\bf B26} (Proc.~Suppl.) (1992) 501;
{\sl Nucl.~Phys.} {\bf B30} (Proc.~Suppl.) (1993) 647.

\bibitem{xue92}
G. Preparata and S-S Xue, `` Emergence of the $\bar tt$-condensate and
the disappearance of scalars in the Standard Model on the Planck Lattice''
MITH 92/21; MITH 93/5.

\bibitem{xue93}
G.~Preparata and S.-S.~Xue, {\sl Phys.~Lett.} {\bf B302} (1993) 442;

\bibitem{planck}
C.W.~Misner, K.S.~Thorne and J.A.~Wheeler, {\it Gravitation\/} (Freeman,
San Fransisco, 1973);\\
J.~Greensite, {\sl Phys.~Lett.} {\bf B255} (1991) 375.

\bibitem{nogo}
H.B.~Nielsen and M.~Ninomiya, {\sl Nucl.~Phys.} {\bf B185} (1981) 20, {\it
ibid.} {\bf B193} (1981) 173, {\sl Phys.~Lett.} {\bf B105} (1981) 219.

\bibitem{nambu}
Y.~Nambu and G.~Jona-Lasinio, {\sl Phys. Rev.} {\bf 122} (1961) 345;\\
W.A.~Bardeen, C.T.~Hill and M.~Linder, {\sl Phys.~Rev.} {\bf D41} (1990) 1647.

\bibitem{tb}
G.~Preparata and S.-S.~Xue, to appear in {\sl Phys.~Lett.} {\bf B} (1994).

\bibitem{bardeen}
W.A.~Bardeen, C.T.~Hill and M.~Linder,
{\sl Phys. Rev. Lett.} {\bf 56} (1986) 1230,
{\sl Nucl. Phys.} {\bf B273} (1986) 649 and {\it ibid.} {\bf B323} (1989)
493,\\
W.A.~Bardeen, C.~T.~Hill and M.~Lindner {\sl Phys.\ Rev.} {\bf D41} (1990)
1647,
\\
W.A.Bardeen, S.~T.~Love and V.A.~Miransky, {\sl Phys.\ Rev.} {\bf D42} (1990)
3514.

\bibitem{xuewill}
G.~Preparata and S.-S.~Xue, in progress.

\end{thebibliography}
\end{document}